\documentstyle[12pt,aaspp4]{article}

\begin{document}

\title{
	Counts and Colors of Faint Galaxies
	in the U and R Bands\altaffilmark{1}
}

\author{
	David W. Hogg,\altaffilmark{2}
	Michael A. Pahre,\altaffilmark{3}
	James K. McCarthy,\altaffilmark{3}
	Judith G. Cohen,\altaffilmark{3} \\
	Roger Blandford,\altaffilmark{2}
	Ian Smail,\altaffilmark{4} \&
	B. T. Soifer\altaffilmark{3}
}

\altaffiltext{1}{
Based on observations obtained at the W. M. Keck Observatory, which is
operated jointly by the California Institute of Technology and the
University of California; and at the Palomar Observatory, which is
owned and operated by the California Institute of Technology.}
\altaffiltext{2}{
Theoretical Astrophysics, California Institute of Technology, Mail
Stop 130-33, Pasadena, CA 91125}
\altaffiltext{3}{
Palomar Observatory, California Institute of Technology, Mail Stop
105-24, Pasadena, CA 91125}
\altaffiltext{4}{
Department of Physics, University of Durham, South Road, Durham
DH1~3LE, Britain}

\begin{abstract}
Ground-based counts and colors of faint galaxies in the $U$ and $R$
bands in one field at high Galactic latitude are presented.
Integrated over flux, a total of $1.2\times 10^5$ sources per square
degree are found to $U=25.5$~mag and $6.3\times 10^5$ sources per
square degree to $R=27$~mag, with $d\log N/dm\sim 0.5$ in the $U$ band
and $d\log N/dm\sim 0.3$ in the $R$ band.  Consistent with these
number-magnitude curves, sources become bluer with increasing
magnitude to median $U-R=0.6$~mag at $24<U<25$~mag and $U-R=1.2$~mag
at $25<R<26$~mag.  Because the Lyman break redshifts into the $U$ band
at $z\approx 3$, at least $1.2\times 10^5$ sources per square degree
must be at redshifts $z<3$.  Measurable $U$-band fluxes of 73~percent
of the $6.3\times 10^5$ sources per square degree suggest that the
majority of these also lie at $z<3$.  These results require an
enormous space density of objects in any cosmological model.
\end{abstract}


\section{Introduction}

The number of faint galaxies as a function of apparent magnitude is
one of the fundamental observational constraints on cosmology and
galaxy evolution.  The first such measurements made with CCDs and
automated source detection algorithms (Hall \& MacKay 1984; Tyson
1988), showed signs of an excess over predictions based on naive
extrapolations of local galaxy properties.  Recent ground-based galaxy
counts to $B=27.5$ (Metcalfe et al 1995), $V=27$, $R=27$ $I=25.5$
(Smail et al 1995), and $K=24$~mag (Djorgovski et al 1995) all reach
integrated number densities around $6\times10^5$~per square degree and
show numbers increasing by a factor of $\sim2$ per magnitude, or
$d\log N/dm\sim 0.3$ in the red and near-infrared.  Given local
luminosity function determinations (e.g., Loveday et al 1992), these
high numbers require either strong source evolution or extreme world
models; it is certainly possible that a number of effects contribute.
Furthermore, the variation of the slope $d\log N/dm$ with observed
waveband, showing increased slope $\sim 0.5$ in the $B$-band and $\sim
0.6$ in the $U$-band and corresponding blueing of the objects with
apparent magnitude (see Koo \& Kron 1992, for a review) is an
important clue to the physical processes generating the radiation in
these objects.

The Hubble Space Telescope (HST) has proven very effective in the
field of faint galaxy counts, benefitting from reduced sky brightness
and the small angular sizes of faint galaxies (Smail et al 1995),
which are only marginally resolved even with HST/WFC's $0.1$~arcsec
resolution.  In a $3\times10^4$~s HST exposure, Cowie et al (1995)
count objects in the field to $I=26$, finding $\approx8\times10^5$ per
square degree.  More recently, in the Hubble Deep Field (Williams et
al 1996), $1.5\times10^5$~s HST exposures were taken in each of four
filters in a single field; counts by Williams et al (1996) find
roughly $10^6$ objects per square degree to
$(F606W)_{AB}\approx30$~mag.

Galaxy counts in the $U$-band have not been pushed to numbers nearly
as high as those in other optical bands.  The subject is interesting
because what $U$-counts do exist show the number counts to rise very
rapidly, by a factor of 3 or 4 per magnitude or $d\log N/dm\sim 0.5$
or $0.6$ (Koo 1986; Songaila et al 1990; Jones et al 1991); i.e.,
$U$-band counts are much steeper than counts at longer wavelengths.
We present the deepest published galaxy counts in the $U$-band,
reaching $U\approx 25.5$~mag, in an image taken under conditions of
good seeing with the Hale Telescope, along with galaxy counts in the
$R$-band, in a deep image taken with the W. M. Keck Telescope.  We
also present $U-R$ colors to look for the color trend implied by the
difference in count slopes.  In terms of point-source sensitivity, the
$U$-band image presented here will be surpassed by ultra-deep HST
observations (such as those in the Hubble Deep Field) but this image
has very good surface-brightness sensitivity and a much wider field of
view.

At $z>3$, the observed $U$ band is at emitted wavelengths shortward of
the Lyman limit, which is expected to be optically thick to absorption
by neutral hydrogen in the intergalactic medium.  This would be
observed as anomalously low $U$-band flux, or anomalously red $(U-B)$
colors for any population of $z>3$ objects.  Guhathakurta et al (1990)
used the lack of such objects to demonstrate that to $R=26$~mag, faint
galaxy counts are not dominated by objects at $z>3$.  Steidel et al
(1995) have counted and, recently, spectroscopically confirmed
(Steidel et al 1996) a population of objects identified for
anomalously red $(U-G)$ colors and find that there are $\sim 1500$
objects in redshift range $3<z<3.4$ per square degree to $R\approx
25$~mag.

\section{Field selection and observations}

The field, RA\,$00^h\,53^m\,23^s\!\!.20$,
Dec\,$+12^{\circ}\,33'\,57''\!\!.5$ (J2000), was chosen for the
purposes of deep $K$-band imaging (Djorgovski et al 1995) and a faint
object redshift survey (Cohen et al 1996a) from among existing deep
Medium Deep Survey (Griffiths et al 1994) HST images which are taken
in parallel mode in fields selected for low extinction and high
Galactic latitude.  The particular MDS field was chosen for its long
HST exposure time and high Galactic latitude to minimize stellar
contamination.  Additional ground-based photometry on this field will
be reported by Pahre et al (1996) and the detailed results of the
redshift survey will be reported by Cohen et al (1996c).

The $U$-band data were taken in 1995 September with the COSMIC
instrument (Dressler 1993) at the prime focus of the 5~m Hale
Telescope.  Palomar CCD13, a Tektronix $2048\times 2048$ array of
$24~\mu$m pixels (TK2048), was used in place of the standard COSMIC
TK2048 CCD owing to high near-UV sensitivity of CCD13.  Individual
$600$~s exposures were taken on a $5\times 5$ grid with roughly
10~arcsec spacing.  The seeing is $1.1$~arcsec in the final stacked
image.  Patchy cloud cover necessitated independent photometric
calibration (see below).

The $R$-band data were taken during periods of fair seeing
($0.8$~arcsec in the final stacked image) on spectroscopic runs in
1995 July and August with the Low Resolution Imaging Spectrograph
(LRIS) (Oke et al 1995) on the 10-m Keck Telescope.  Individual
$600$~s exposures were taken at dithers of several arcseconds with
respect to one another.  A temporary problem with the telescope caused
the individual $R$-band exposures to be contaminated with transient,
non-repeating streaks of scattered light, from bright stars outside
the field reflecting from telescope structures.  This limits the
quality of the $R$ image flatfield and background estimation, and
reduces the completeness and quality of the photometry at very faint
levels.  In removing the streaks, small gradients on scales of $\sim
20$~arcsec were eliminated (taking with them any hypothetical
population of large, low-surface-brightness galaxies; of course the
detection algorithm is not optimized for such objects anyway).  Again
patchy cloud cover necessitated independent photometric calibration.

For both the $U$-band and $R$-band data, before stacking, individual
images were shifted and geometrically remapped to a Cartesian plane
according to a distortion map determined from the dithered images
themselves.  The remapped, shifted images were stacked with the {\em
IRAF/imcombine\/} task, making use of sigma-clipping to remove cosmic
rays.  Matched $1\times1~{\rm arcmin^2}$ sections of the stacked Hale
$U$ and Keck $R$-band images are shown in Figs.~1 and~2, and
observational details are summarized in Table~1.
\begin{deluxetable}{clccccc}
\tablewidth{0in}
\tablecaption{Imaging data}
\tablehead{
\colhead{band}
 & \colhead{instrument}
 & \colhead{field area}
 & \colhead{total $t_{\rm exp}$}
 & \colhead{pixel scale}
 & \colhead{seeing\tablenotemark{a}}
 & \colhead{$3\,\sigma$ detection\tablenotemark{b}} \\
 &
 & \colhead{(arcmin$^2$)}
 & \colhead{(s)}
 & \colhead{(arcsec)}
 & \colhead{(arcsec)}
 & \colhead{(mag)}
}
\startdata
$U_{13}$    & Hale/COSMIC/CCD13 &  81 & 28000 & 0.28 & 1.1 & 26.36 \\
$U$         & P60/CCD13         & 160 &  5400 & 0.37 & 2.2 & 23.13 \\
$R$         & Keck/LRIS         &  39 &  8400 & 0.22 & 0.8 & 28.01 \\
$R_{\rm C}$ & P60/CCD13         & 160 &   600 & 0.37 & 1.3 & 23.25 \\
\enddata
\tablenotetext{a}{
Seeing FWHM in final, stacked images}
\tablenotetext{b}{
Signal-to-noise ratio of $3$ through an aperture of diameter $1.5$
times the seeing FWHM, with no aperture correction applied}
\end{deluxetable}
\begin{figure}
\figurenum{1}
\plotone{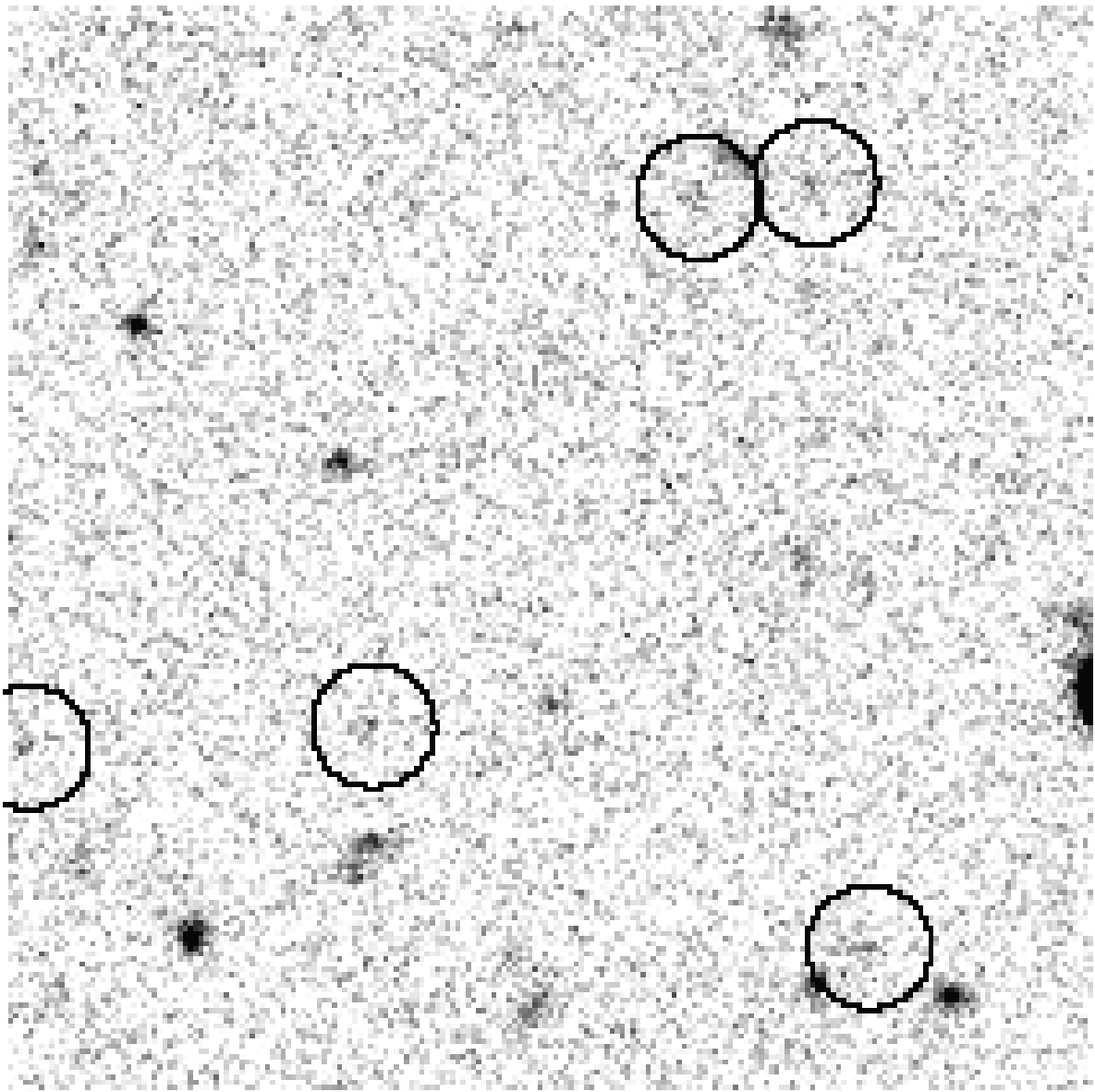}
\caption{A $1\times 1$~arcmin$^2$ section of the final, stacked
$U$-band image.  Sources with $25.0<U_{13}<25.5$~mag are circled.}
\end{figure}
\begin{figure}
\figurenum{2}
\plotone{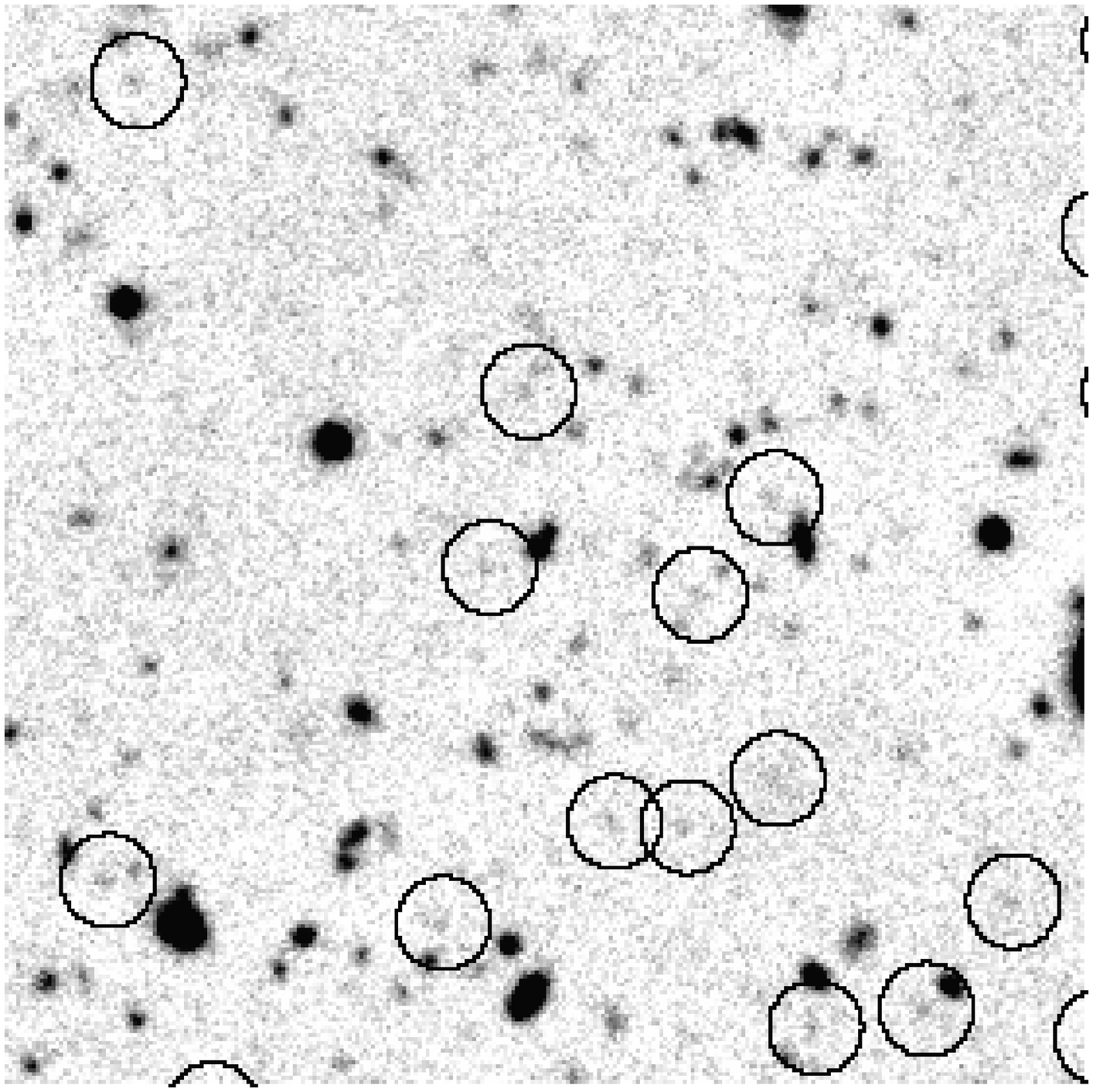}
\caption{A $1\times 1$~arcmin$^2$ section of the final, stacked
$R$-band image, matched to Fig.~1.  Sources with $26.5<R<27.0$~mag are
circled.}
\end{figure}

The $U$ and $R$-band images were calibrated with $U$ and $R_{\rm
C}$-band CCD images taken under photometric conditions with CCD13 on
the Palomar Sixty-Inch Telescope (P60).  The details of the P60 data
reduction are presented in Pahre et al (1996).  Calibration shows that
the filter and CCD used on the Hale Telescope make a non-standard $U$
bandpass, which we denote ``$U_{13}$''.  We find
$U=U_{13}+0.06\,(U_{13}-R)$.  In the following we use $U_{13}$ but the
magnitudes can be corrected to pseudo-$U$-band by adding $0.08$~mag,
correct for a typical faint object, which has $(U_{13}-R)=1.3$~mag.
This correction is on the same order as the estimated uncertainty in
the calibration, between $0.05$ and $0.1$~mag.

\section{Analysis}

\subsection{Object detection}

Objects are detected in both the $U$ and $R$-band images with the
image analysis package {\em SExtractor\/} (Bertin \& Arnouts 1996).
The algorithm is: (1)~fit a smooth surface to the background;
(2)~convolve the image with a Gaussian filter with FWHM matched to the
PSF; and (3)~find objects above the threshold which corresponds to a
point source with $U_{13}=25.63~{\rm mag}$ or $R=28.49~{\rm mag}$.
Often detected objects have multiple peaks; a peak is split off into
its own object only if its part of the object contains $>10$~percent
of the total flux.  We did not use the package for star/galaxy
separation or any ``cleaning'' of spurious detections.  Stars only
make up a small fraction of the faint sources at high Galactic
latitude (Smail et al 1995) and cleaning spurious objects is handled
by our noise object and completeness corrections described below.

\subsection{Photometry}

In order to match the seeing of our images ($1.1$~arcsec FWHM in the
$U$-band and $0.8$~arcsec in the $R$-band), aperture photometry is
performed in $1.5~{\rm arcsec}$ diameter apertures, which is between
1.3 and 2 times the seeing FWHM.  To these aperture magnitudes
aperture corrections are added to attempt, in a statistical way, to
account for object flux coming from outside the aperture.  The
faintest objects are consistent with having stellar profiles, in these
images and even in $R$-band images with superior seeing (Smail et al
1995), so we added the stellar correction, $-0.60$~mag in the $U$-band
image and $-0.25$~mag in the $R$-band.  For all objects we also
measure a $1\sigma$ isophotal magnitude to which is added an aperture
correction varying linearly from $0$ to $-0.60$ over the range
$23.5<U_{13}<25.5$~mag and from $0$ to $-0.25$ over $24.5<R<26.0$~mag.
Every object is assigned the brighter of the two corrected magnitudes.

The reddening in this field is $A_V=0.13$~mag (Burstein \& Heiles
1982) so we corrected the $U$-band fluxes by $0.20$~mag and $R$-band
by $0.11$~mag.

\subsection{Removal of noise objects}

The transforming, shifting and stacking of the images introduces
pixel-to-pixel correlations in the noise, making it difficult to
estimate analytically the contamination of the counts by spurious
detections of peaks in the noise.  Spurious detections of this type
are corrected-for in a statistical way, by subtracting ``noise
counts'' from the positive counts.  Noise counts are estimated by
running the detection and photometry algorithms on the images but
searching for negative rather than positive objects.  The raw negative
counts are subtracted from the raw positive counts before applying
completeness corrections (see below) and all Poisson error bars
include the uncertainty added by this procedure.  In the $U$-band
image, noise objects account for $10$~and $29$~percent of the
objects in the two faintest half-magnitude bins and in the $R$-band
image, only $0.6$~and $8$~percent in the two faintest half-magnitude
bins.

Spurious objects are sometimes produced by cosmic ray hits.  However,
so many individual exposures (47 in $U_{13}$, 14 in $R$) were stacked
with the sigma-clipping algorithm that no significant-flux cosmic-ray
events could plausibly remain.

\subsection{Completeness correction}

The counts are corrected for completeness in a manner similar to that
of Smail et al (1995).  Detected objects are cut out of the final $U$
and $R$-band images, dimmed by a factor of $10$, and added back into
randomly selected subfields.  The detection algorithm is run and the
catalogs of objects in the subfields are compared before and after
adding the additional object.  By this procedure we generate a
``completeness matrix'' $P_{ij}$ which stores the probability that an
added object of true magnitude $m_i$ is in fact detected with
magnitude $m_j$.  Because some objects are not recovered at all and
because others are blended into existing brighter objects, etc., the
sum over $j$ of $P_{ij}$ will not in general be unity.  The
completeness information is a matrix because the statistical
incompleteness in each magnitude bin depends on the true functional
form of the number-magnitude plot; however, under the assumption that
the galaxy count slopes do not change dramatically over the magnitude
range of interest, this matrix converts naturally into a fractional
completeness as a function of magnitude.

To construct $P_{ij}$ in practice, bright objects were cut out of the
images, dimmed by a factor of 10, replaced at random locations, and
then searched-for and photometered by the detection algorithm.  In the
$U$-band image, $10^5$ random replacements were performed, and in the
$R$-band image, $6\times10^4$, with the number increasing with
magnitude in proportion to the counts.  Each element of the matrix
also has an associated uncertainty from Poisson statistics.  The
detection fraction $f_j$ in each bin was generated by assuming that
the $U$ counts follow a power law $d\log N/dm\approx 0.47$ and that
the $R$ counts follow power law $d\log N/dm\approx 0.33$.  Our
noise-object-subtracted, completeness-corrected $U$ and $R$-band
object counts are shown in Figs.~3 and~4, along with the
completeness-corrected counts of other authors.  Note that because
this completeness correction corrects not just for missing numbers but
also for photometry errors, and because the counts increase with
magnitude, it is possible for the completeness correction near (but
not at) the detection threshold to be negative, as is seen for at
least one point in Fig.~3.
\begin{figure}
\figurenum{3}
\plotone{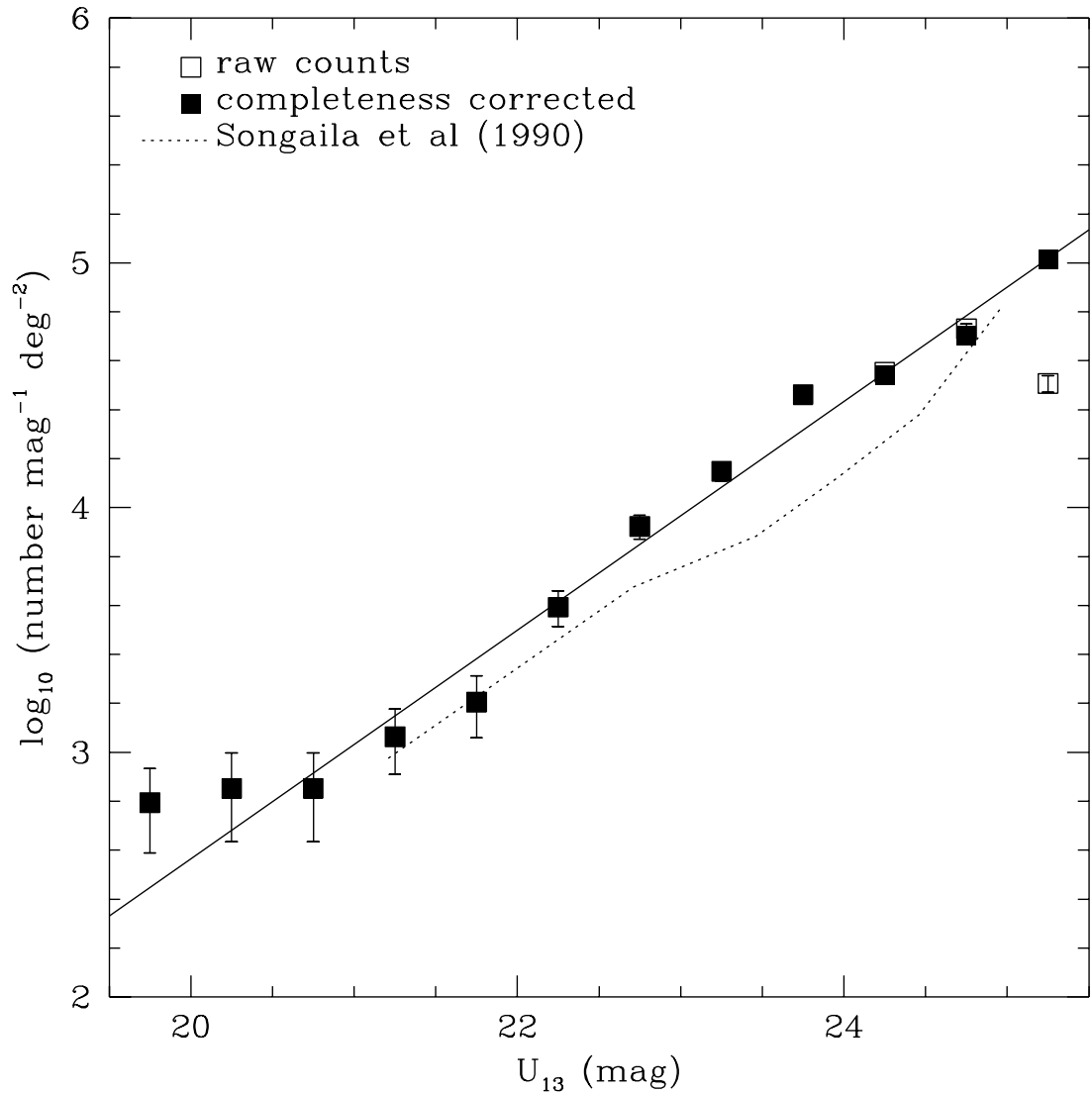}
\caption{
The $U_{13}$-band number counts.  Open squares are the positive counts
with the ``noise'' counts (negative counts, see text) subtracted, from
the data presented here; error bars are Poisson noise for the
difference.  Filled squares are the completeness-corrected counts (see
text), also from the data presented here.  The solid line is a fit to
the corrected counts; it has slope $d\log N/dm=0.467$.  The dotted
line shows counts of Songaila et al (1990), converted to this
magnitude system assuming $U_{13}\approx U'_{AB}-0.79$~mag for a
typical object.}
\end{figure}
\begin{figure}
\figurenum{4}
\plotone{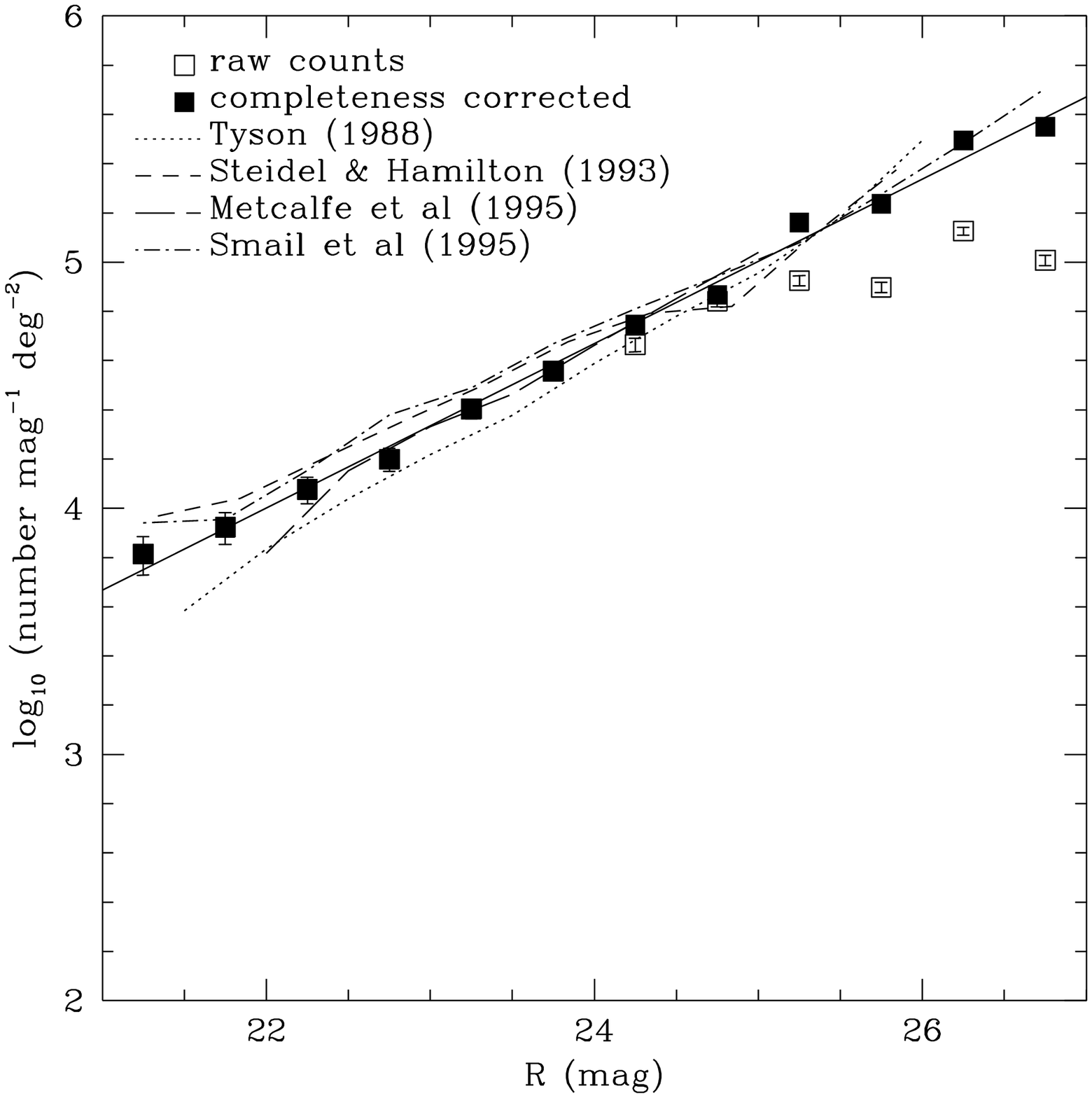}
\caption{
The $R$-band number counts.  Open squares are the positive counts with
the ``noise'' counts (negative counts, see text) subtracted, from the
data presented here; error bars are Poisson noise for the difference.
Filled squares are the completeness-corrected counts (see text), also
from the data presented here.  The solid line is a fit to the
corrected counts; it has slope $d\log N/dm=0.334$.  Dashed and dotted
lines are the completeness-corrected counts of other authors.  The
Steidel \& Hamilton (1993) counts have been converted to this
magnitude system assuming $R\approx{\cal R}_{AB}-0.16$ for a typical
object.}
\end{figure}

Faint objects get smaller in angular size with increasing magnitude
(Smail et al 1995; Im et al 1995), and since more compact objects are
easier to detect at the same flux level, the detected fractions
calculated by this technique would, in better data, be lower limits.
A better procedure would involve changing the angular sizes of the
objects as well as dimming them before replacing them.  However, the
seeing in our images is not good enough for the changing sizes to be a
significant effect at faint levels.

\subsection{Color measurement}

Magnitudes through $1.5$~arcsec diameter apertures, measured with the
NOAO {\em apphot\/} package, were subtracted to make $(U-R)$ colors
for the entire $R$-selected sample to $R=27$ and that part of the
$U$-selected sample which overlaps the $R$-band image to $U=25$.
Median colors as a function of magnitude and color histograms for
several $U$-selected and $R$-selected subsamples are shown in
Figure~5.
\begin{figure}
\figurenum{5}
\plotone{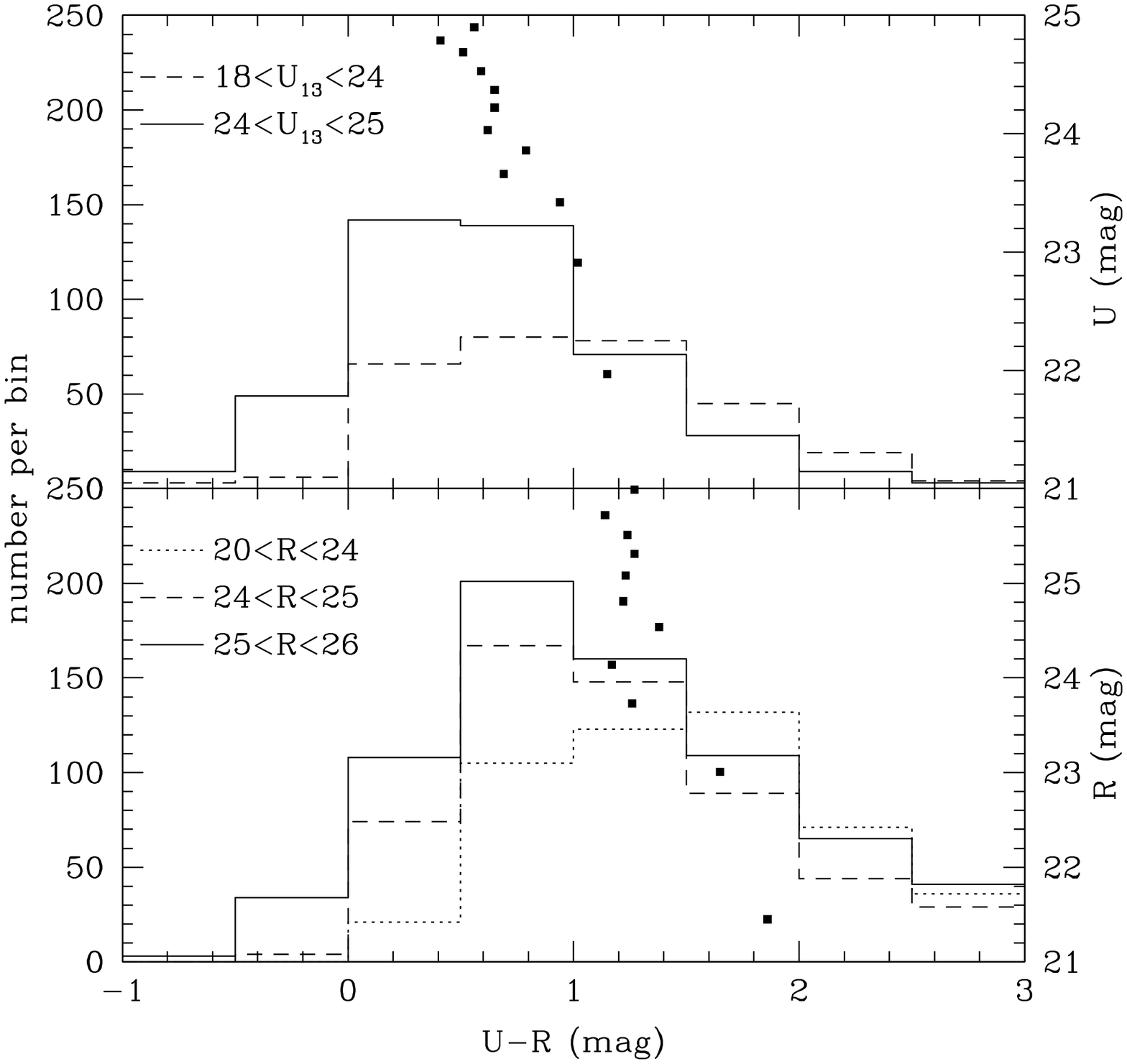}
\caption{
Histograms of $U-R$ colors (corrected to the standard $U$-band)
through $1.5$~arcsec apertures for several $U$-selected (top) and
$R$-selected (bottom) samples.  The squares show median colors as a
function of magnitude, medianed in groups of 61 in the $U$-band and
groups of 201 in the $R$-band.}
\end{figure}
To $R=27$~mag, 73~percent of the sources in the $R$-band image are
detected with a confidence of $1\,\sigma$ or better in the $U$-band
image.  The median measured color in the faintest $U$-band magnitude
bin is $U-R=0.6$~mag.  In the faintest $R$-band magnitude bin it is
$U-R=1.3$~mag, although clearly there are large magnitude
uncertainties in the $U$-band data for such faint objects.

\section{Results and discussion}

\subsection{Colors and faint-end count slopes}

The faint ends of the counts have different slopes in the $U$ and $R$
bands, which means, under fairly robust assumptions, the mean object
color must be a strong function of magnitude, a trend clearly visible
in the color-magnitude diagram.  Since the night sky would be
infinitely bright in the $U$ band if this trend continued forever, at
some faint magnitude the $U$-band count slope must break to match the
$R$-band slope and thereby end the bluing trend.  One physical
explanation for the bluing trend is that the fraction of the radiation
due to young stars may increase with apparent magnitude.  In this
scenario, the objects should get no bluer than about
$f_{\nu}\propto\nu^0$, the ultraviolet spectral slope of a
star-forming galaxy (Kinney et al 1993), so a natural prediction is
that the break in the $U$-band counts will appear when the median
$(U-R)$ obtains the $f_{\nu}\propto\nu^0$ value, $-0.4$~mag.  This
should happen at $27<U<28$~mag.  Very deep counts in the Hubble Deep
Field data (Williams et al 1996), which reach point-source detection
levels several magnitudes fainter than these observations, report a
break at $U\sim25.3$, although it is not robustly detected.

As is shown in Figures~3 and~4, the $U$-band counts found here are in
good agreement with the previous study near this depth (Songaila et al
1990), the $R$-band counts are also in very good agreement with
previous studies (Smail et al 1995; Metcalfe et al 1995; Steidel \&
Hamilton 1993; Tyson 1988) as are the colors (Guhathakurta et al 1990;
Steidel \& Hamilton 1993).  The results here are similar to those of
Guhathakurta et al (1990) except that the latter have $<2.7\times
10^5$ sources per square degree and only identify sources to $R\approx
26$, although that limit is fuzzy because sources were identified by
those authors not in a single band but in a summed $UBR$ image.

\subsection{The number problem}

We count faint objects in the $U$ and $R$-bands to surface densities,
integrated over flux, of $1.2\times 10^5$ per square degree to
$U=25.5$~mag and $6.3\times 10^5$ per square degree to $R=27$~mag.
These numbers correspond to $0.5$ and $2.5\times10^{10}$ objects over
the entire sky.  These numbers are in excess of the total number of
galaxies expected within the observable Universe in any no-evolution
or passive-evolution model (i.e., models in which the comoving number
density of galaxies is conserved), for the following simple reason: If
the local galaxy luminosity function is that given by Loveday et al
(1992) (or Mobasher et al 1993; or Lin et al 1996) and galaxies exist
down to luminosities $5$~mag fainter than $L^{\ast}$, the space
density of galaxies is $5.7\times 10^{-2}\,h^3~{\rm Mpc^{-3}}$, where
$h$ is the Hubble constant in units of $100~{\rm
km\,s^{-1}\,Mpc^{-1}}$.  In units of Hubble volume $V_H\equiv
(c/H_0)^3$, this corresponds to a density of $1.5\times
10^9\,V_H^{-1}$.  This number density increases by only $3.5\times
10^8\,V_H^{-1}$ per magnitude as the luminosity function is
extrapolated at the faint end.  In an Einstein-de~Sitter universe,
$(\Omega_M,\Omega_{\Lambda})=(1,0)$, there is an all-sky comoving
volume of only $4.2\,V_H$ to $z=3$.  The product of number density and
volume is $6\times 10^9$ galaxies of the entire sky, so there are 4
times too many sources observed in these data to be easily explained
by naive models.  The number of galaxies increases with apparent
magnitude, so the problem is worse with deeper data: Cowie et al
(1996) show a factor of 10 times too many sources, and the counts in
the HDF (Williams et al 1996) show a factor of at least 15 times too
many sources.  We know that only a small fraction of the galaxies
presented here lie beyond $z\approx3$ because at this redshift the
Lyman limit redshifts into the $U$-band and the flux is cut off by
either self-absorption or absorption in intervening material (e.g.,
Guhathakurta et al 1990; Steidel et al 1995, 1996); 73~percent of the
sources to $R=27$~mag are detected in $U_{13}$.  Of course these
numbers are consistent with the population of galaxies at $z>3$ found
and confirmed spectroscopically by Steidel et al (1995; 1996), because
that high redshift population is only a few percent of the total
source counts.

For $(\Omega_M,\Omega_{\Lambda})=(0.05,0)$ the comoving volume is
$14\,V_H$, for $(\Omega_M,\Omega_{\Lambda})=(0.2,0.8)$ it is
$19\,V_H$, so switching world models does not solve the problem unless
one considers even more extreme world models that are almost certainly
ruled out by gravitational lens statistics (Turner 1990).  Several
authors have found local luminosity function amplitudes higher than
that of Loveday et al (1992) by factors of a few (Marzke et al 1994;
Lilly et al 1995; Small 1996) but factors large enough to solve the
number problem would be surprising.  There is also some disagreement
over the faint-end slope $\alpha$ (in the parameterization
$\phi(L)\propto L^{\alpha}$) of the local luminosity function (e.g.,
Lilly et al 1995).  If the slope is steeper than the standard flat
value (i.e., if $\alpha$ is more negative than $\alpha=-1.0$), there
could be significantly more galaxies in the local Universe than the
above estimates suggest.  For example, comparing with $\alpha=-1.0$,
if $\alpha=-1.5$ but $L^{\ast}$ and the bright-end amplitude are held
fixed, there is a factor of $\approx 5$ more galaxies to a luminosity
limit of $5$~mag fainter than $L^{\ast}$.  This would go some way
towards alleviating the number problem, although the slope is
well-enough determined by local surveys that such a large discrepancy
seems unlikely.

With current observational constraints, it is more natural to look to
evolution in the sources to solve the number problem.  It has been
suggested that the galaxy merging rate is high, so large numbers of
small galaxies at high redshift evolve into small numbers of large
galaxies locally (Guiderdone \& Rocca-Volmerange 1991; Broadhurst et
al 1992).  It is possible that there is a large population of small
galaxies which explode or evaporate after their first burst of star
formation and supernovae (Babul \& Rees 1992).  Also, there is a
less-explored possibility that galaxies may form at high redshift with
more-or-less their present-day masses but if star formation occurs in
small, spatially isolated bursts (Katz 1992), each present-day galaxy
would be observed as many different objects at large lookback time.
There may be support for some of these models in the redshift
distribution of faint galaxies; several authors have found a
steepening of the faint-end slope of the luminosity function with
redshift (Eales 1993; Lilly et al 1995; Ellis et al 1996), and
enormous information will come from the current generation of
super-deep redshift surveys (e.g.  Cowie et al 1996; Cohen et al
1996a, 1996b; Koo et al 1996).

\subsection{Stronger redshift limits with HST}

The ultraviolet sensitivities of the WFPC2 and STIS instruments (the
latter to be installed in 1997) on HST suggest the extension of this
technique---limiting redshift distributions by looking for the Lyman
limit break---to lower redshift.  In fact, the broadband UV filters of
these instruments can actually be used to locate the Lyman break over
a range of redshifts from 0.9 to 1.8, a range which is difficult to
identify at present even with 10-m--class telescopes because there are
very few lines in the useful window of ground-based visual
spectrocopy.  With a modest amount of observing time it will be
possible to obtain at least statistical redshift distributions
significantly deeper than the practical limits of ground-based
spectroscopy even if there were spectroscopic features in this
redshift range.  Such observations are particularly crucial since many
of the above-mentioned models make very different predictions for the
fraction of galaxies at $z>1$ and $z>2$.

\acknowledgements
We are grateful to the W. M. Keck Foundation for the vision to fund
the construction of the W. M. Keck Observatory.  We benefited from
helpful conversations with Kurt Adelberger, John Gizis, Tomislav
Kundi\'c, Gerry Neugebauer, Sterl Phinney, and Chuck Steidel, and from
comments from Richard Ellis and an anonymous referee.  We are grateful
for financial support from NSF grant AST 92-23370 (DWH, RB), NSF grant
AST-91-57412 and the Bressler Foundation (MAP), support through a
PPARC Advanced Fellowship (IS), and grants from NSF and NASA (BTS).
The calibrated images used for this investigation will be made
available to the public; contact the authors for more information.

\end{document}